\documentclass[11pt]{article}
\pdfoutput=1

\usepackage{euscript}
\usepackage{amssymb}
\usepackage{amsfonts}
\usepackage{amsbsy}
\usepackage{epsfig}
\usepackage{amsthm}
\usepackage{amscd}
\usepackage{amstext}
\usepackage{verbatim}
\usepackage{amsmath}
\usepackage{cancel}
\usepackage{capt-of}
\usepackage{empheq}
\usepackage{subfigure}

\usepackage[pdftex]{hyperref}

\def\ben{\begin{equation}}
\def\een{\end{equation}}
\def\half{{\textstyle{1\over2}}}

   \let\k=\kappa

\def\be{\begin{equation}}
\def\ee{\end{equation}}
\def\beq{\begin{equation}}
\def\eeq{\end{equation}}
\def\ba{\begin{array}}
\def\ea{\end{array}}

\def\dalemb#1#2{{\vbox{\hrule height .#2pt
       \hbox{\vrule width.#2pt height#1pt \kern#1pt
               \vrule width.#2pt}
       \hrule height.#2pt}}}

\newcommand{\bea}{\begin{eqnarray}}
\newcommand{\eea}{\end{eqnarray}}
\newcommand{\tr}{{\rm tr} }

\newcommand{\dd}{{\rm d}}

\renewcommand{\eqref}[1]{(\ref{eq:#1})}
\renewcommand{\Bar}[1]{\overline{#1}}

\def\ocal{{\mathcal{O}}}

\begin{document}

\begin{center}

{ \Large {\bf
Emergent scale invariance of disordered horizons
}}

\vspace{1cm}

Sean A. Hartnoll$^{1}$, David M. Ramirez$^{1}$  and Jorge E. Santos$^{2}$

\vspace{1cm}

{\small
$^{1}${\it Department of Physics, Stanford University, \\
Stanford, CA 94305-4060, USA }}

\vspace{0.5cm}

{\small
$^{2}${\it Department of Applied Mathematics and Theoretical Physics, \\
University of Cambridge, Wilberforce Road, \\
Cambridge CB3 0WA, UK}}

\vspace{1.6cm}

\end{center}

\begin{abstract}

We construct planar black hole solutions in AdS$_3$ and AdS$_4$
in which the boundary CFT is perturbed by marginally relevant quenched
disorder. We show that the entropy density
of the horizon has the scaling temperature dependence $s \sim T^{(d-1)/z}$ (with $d=2,3$).
The dynamical critical exponent $z$ is computed numerically and, at weak disorder, analytically.
These results lend support to the claim that the perturbed CFT flows to a disordered quantum
critical theory in the IR.

\end{abstract}

\pagebreak
\setcounter{page}{1}

\section{Introduction}

Consider a quantum field theory in the presence of quenched disorder, that is, spatially random couplings.
As with simpler spatially uniform couplings, the effects of quenched disorder can be relevant or 
irrelevant \cite{harris, sachdev}. When relevant, the disorder can drive the quantum theory
into various possible nontrivial low energy phases. Of interest to us in this paper will be disordered fixed points.
At a disordered fixed point, the low energy physics exhibits an emergent scale invariance.
Such fixed points have a different structure to, for example, relativistic conformal field theories because
the spatial dependence of the couplings at the fixed point means that momentum is not conserved at long wavelengths.

It is not easy to find controlled instances of disordered fixed points where the disordered coupling
is stabilized at a finite value \cite{sachdev}. Nonetheless, such critical theories without momentum conservation are very interesting
candidates to understand the universal behavior of bad metals \cite{Hartnoll:2014lpa}. It is therefore of interest
to have concrete examples at hand. Evidence for a disordered fixed point was recently found in a holographic system \cite{Hartnoll:2014cua}.
The disordered fixed point itself is dual to a highly inhomogeneous extremal black hole horizon.

In \cite{Hartnoll:2014cua}, a CFT described by a gravity dual was perturbed by marginally relevant disorder.
It was shown that the {\it disorder averaged} geometry in the far IR regime exhibited an emergent scaling invariance
\be\label{eq:IR}
\overline{ds^2} \; \stackrel{r\to\infty}{=} \; L^2_\text{IR} \left(- \frac{dt^2}{r^{2z}} + \frac{dr^2 + d\vec x^{\, 2}}{r^2} \right) \,.
\ee
Here $\overline{ds^2}$ is the disorder averaged metric. This result was obtained both by resummation of
perturbation theory and by full-blown numerics. The averaged metric was therefore seen to be characterized by
a dynamical critical exponent $z$ that determines the relative scaling of space and time \cite{Kachru:2008yh}.
However, it was not completely clear what physical quantities would be determined by this disorder averaged metric.

In this paper we will revisit the system considered in  \cite{Hartnoll:2014cua}, but now placed at a nonzero temperature.
Our main result, that we again obtain both analytically and numerically, is that the entropy density of the system scales
with temperature as
\be\label{eq:avS}
s  \sim T^{(d-1)/z} \,,
\ee
with $d$ the number of spacetime dimensions of the quantum field theory and
with the same $z$ appearing as in the disorder averaged zero temperature IR metric (\ref{eq:IR}).
Namely, in an expansion in small disorder strength $ {\bar V}$,
\be
z = 1 + \half \pi^{d/2-1} \Gamma\left({\textstyle  \frac{d}{2} } \right) {\bar V}^2 + \ocal \left({\bar V}^4 \right) \,.
\ee
Thus, we have shown that the averaged metric indeed accurately captures the scaling properties of a disordered fixed point.
In fact, we will see that the temperature scaling of the entropy (\ref{eq:avS}) is equal to the temperature scaling of the entropy of the averaged metric, although the coefficients need not agree. 

The main technical achievements in this paper are the perturbative and numerical construction of the $T>0$ disordered black hole spacetimes. \S\ref{sec:setup} describes the general setup. \S\ref{sec:pert} and \S\ref{sec:therm} obtain the solution perturbatively in disorder and describe a resummation of logarithms. \S\ref{sec:num} constructs the solutions numerically. In \S\ref{sec:diss} we discuss open questions.

\section{Setup}
\label{sec:setup}

In this section we review the holographic description of a CFT perturbed by marginal disorder
 \cite{Hartnoll:2014cua}. The starting point is a real scalar coupled to gravity in $d+1$
dimensions.\footnote{Throughout our discussion we will let $d$ denote
  the boundary spacetime dimensions. We will always work in signature
  $(-++\dotsb+)$. For index conventions, capital Latin indices will
  denote all bulk directions, lowercase Latin letters, $a,b$ etc.~,
  will denote boundary spacetime directions, while middle lowercase
  Latin letters, $i,j$ etc.~, will denote the boundary spatial
  directions.} The action is:
\begin{align}
  S ={}& \frac{1}{2 \kappa_N^2} \int \dd^{d+1} x \, \sqrt{-g} \left[ R -
    \Lambda - 2 \nabla_A \Phi \nabla^A \Phi - 4 V(\Phi) \right]\,
  . \label{eq:action}
\end{align}
Here $\kappa_N^2 = 8 \pi G_N$ and $\Lambda = - \frac{d(d-1)}{L^2}$ is
the usual AdS$_{d+1}$ cosmological constant. The resulting equations
of motion are:
\begin{align}
  0 ={}& \Box \Phi - V'(\Phi)\, , & R_{AB} ={}& 2 \nabla_A \Phi
  \nabla_B \Phi + \frac{1}{d-1} g_{AB} \left[ 4V(\Phi) + \Lambda
  \right]\, .
  \label{eq:einstein}
\end{align}

For the scalar potential, we take a negative mass squared:
\begin{align}
  V(\Phi) ={}& - \frac{\mu}{2 L^2} \Phi^2\, . \label{eq:vphi}
\end{align}
The holographic dictionary (e.g. \cite{Hartnoll:2009sz}) tells us that $\Phi$ is dual to an operator
${\cal O}$ with dimension:
\begin{align}
  \Delta ={}& \frac{d}{2} + \sqrt{ \left(\frac{d}{2}\right)^2 - \mu}\,
  . \label{eq:dimension}
\end{align}
More explicitly, this is seen by considering the asymptotic behavior
of the scalar near the AdS$_{d+1}$ boundary. In the Poincar\'e patch,
where the line element approaches:
\begin{align}
  \dd s^2 ={}& \frac{L^2}{r^2} \left( \eta_{ab} \dd x^a \dd x^b + \dd
    r^2 + \dotsb \right)\, ,
    \label{eq:fgcoordinates}
\end{align}
as $r \to 0$, the scalar has the following form near the boundary:
\begin{align}
  \Phi(r \to 0) ={}& r^{d-\Delta} \Phi_1(x^a) + r^{\Delta} \Phi_2(x^a) +
  \dotsb\, . \label{eq:bdyscalar}
\end{align}
The correspondence then tells us that $\Phi_1$ is identified with the
source for ${\cal O}$ while $\Phi_2$ encodes the response \cite{Hartnoll:2009sz}. 

Our interest in this paper is to consider the effect of a 
disordered source for ${\cal O}$ at finite temperature $T$. To be
explicit, we will work with a short ranged, quenched, Gaussian
disorder ensemble, where the ensemble of sources is determined
by:
\begin{align}
  \Bar{\Phi_1(x^i)} ={}& 0\, , & \Bar{\Phi_1(x^i) \Phi_1(y^i)} ={}&
  \bar V^2 \delta^{(d-1)}(x^i - y^i)\, . \label{eq:disorder-dist}
\end{align}
All other moments of the distribution are then fixed by Wick
contraction. Note that, as befits quenched disorder, the random
sources only depend on the boundary spatial directions. Our analytic
discussion later will involve a resummation of perturbation theory in
$\bar V$, whereas the numerics will be exact in $\bar V$.

We will focus on the case of `marginal' disorder;
that is, we will take the distribution to saturate the Harris
criterion, which determines when short-range disorder affects critical
phenomena \cite{harris, sachdev}. A simple heuristic way to see this result is to note that
since $\Phi$ is dual to an operator of dimension $\Delta$, dimensional
analysis tells us that $[\Phi_1] = d - \Delta$ and therefore
\eqref{disorder-dist} suggests we assign $\bar V$ a dimension of:
\begin{align}
  2 [\Phi_1] = 2 [\bar V] + d-1 \qquad \Rightarrow \qquad [\bar V]
  ={}& \frac{d+1}{2} - \Delta \, . \label{eq:vbar-dim}
\end{align}
We expect then that the disorder is relevant if $\Delta <
\frac{d+1}{2}$, irrelevant for $\Delta > \frac{d+1}{2}$ and marginal
for $\Delta = \frac{d+1}{2}$. Requiring $[\bar V] = 0$ fixes the value
of $\mu$ in (\ref{eq:dimension}) to be:
\begin{align}
  \mu ={}& \frac{d^2-1}{4}\, .
  \label{eq:harris}
\end{align}

To realize the disorder, we will use a spectral representation \cite{Shinozuka1991},
writing the source as:
\begin{align}
  \Phi_1(x^i) ={}& \bar V \sum_{\{n_i\}=1}^{N-1} C_{\{n_i\}}
  \prod_{i=1}^{d-1} \cos (k_{i,n_i} x^i + \gamma_{n_i})\,
  . \label{eq:bdy-source}
\end{align}
Here the $\gamma_{n_i}$ are random phases uniformly distributed over
$(0,2\pi)$ while the specifics of the distribution are determined by
the constants $C_{\{n_i\}}$ and the selection of $k_{i,n_i}$. To
strictly capture the disorder in the thermodynamic limit we must take
$N \to \infty$. The disorder average of a quantity $f$ is then given
by:
\begin{align}
  \Bar{f} ={}& \lim_{N \to \infty} \int
  \left[\prod_{i=1}^{d-1}\prod_{n_i=1}^{N-1} \frac{\dd
      \gamma_{i,n_i}}{2\pi} \right]\, f\, .
\end{align}
We will consider the simplest case of a short range, Gaussian and isotropic disorder distribution, which
corresponds to:
\begin{align}
  C_{\{n_i\}} ={}& C = \left(2 \sqrt{\Delta k}\right)^{d-1}\, , &
  k_{i,n_i} ={}& n_i \Delta k\, , & \Delta k ={}& \frac{k_0}{N} \,
  . \label{eq:gauss-dist}
\end{align}
The wavevectors of the modes making up the disordered source (\ref{eq:bdy-source})
therefore range from $k_0/N$ to $k_0$. These are the IR
and UV cutoffs on the disorder distribution, respectively.
In principle we could take the spacings $\Delta k$ to depend on
the direction of $k$, but for simplicity we take an isotropic
distribution.

Since we will be working at finite temperature, it is important to
keep the various scales in mind. It is useful to consider the two
dimensionless parameters: $\kappa_0 = k_0/T$ and
$\kappa_{\text{IR}} = \kappa_0/N$. The spectral representation
requires we take $N \to \infty$ and physically we want $k_0 \gg T$,
but the order of limits is important. Since our aim is to describe a
disordered system at small but finite temperature, we should be taking
the $N \to \infty$ limit first, and so in what follows we will work
with the following hierarchy:
\begin{align}
  \kappa_\text{IR} \ll 1 \ll \kappa_0\, , \qquad \text{i.e.} \qquad \frac{k_0}{N} \ll T \ll k_0 \,.
  \label{eq:hierarchy}
\end{align}

\section{Perturbative geometry}
\label{sec:pert}

In this section we perturbatively
construct the spacetime deformed by the disordered
boundary source (\ref{eq:gauss-dist}). This involves solving the bulk
scalar field wave equation subject to the disordered boundary
condition, computing the energy momentum tensor of this scalar field, and
then computing the backreaction on the metric.
Our analytic discussion will largely focus on the lowest order thermodynamic
corrections. These will be logarithmic in temperature, suggesting a natural
resummation. We will show later in section \ref{sec:config} that to obtain
the entropy as a function of temperature to this lowest order, it is sufficient to
obtain the metric that follows from the backreaction
produced by the \emph{disorder averaged} scalar stress tensor. This amounts to finding
the leading disorder averaged correction to the metric, which we will now do.

In the appendix we specialize to the $d=2$ dimensional case and determine the correction to the
geometry for generic scalar configurations, without averaging. It will be noted that despite
the expressions being rather complicated, no essentially new physics is found from the full, configuration dependent expressions.

\subsection{Geometry at ${\cal O}(\bar V^0)$}

We will work throughout in the Poincar\'e patch at finite
temperature. Therefore in the limit $\bar V \to 0$, the line element
reduces to (from here on we set $L=1$):
\begin{align}
  \dd s^2 ={}& \frac{1}{r^2} \left[ - f(r) \dd t^2 + \frac{\dd
      r^2}{f(r)} + \sum_{i=1}^{d-1} (\dd x^i)^2 \right]\,
  . \label{eq:back-geo}
\end{align}
Here $f(r)$ is the emblackening factor $f(r) = 1 - (\frac{r}{r_+})^d$,
where $r_+$ is the horizon radius. In terms of the temperature $T$, we
have $T = \frac{d}{4 \pi r_+}$. The entropy density of the
thermal state in the dual field theory is then given by the familiar
Bekenstein-Hawking entropy:
\begin{align}
  s ={}& \frac{1}{4 G_N} \frac{1}{V} \int \dd^{d-1} x^i \sqrt{- \gamma} =
  \frac{1}{4 G_N} \frac{1}{r_+^{d-1}} = \frac{1}{4 G_N} (4\pi)^{d-1} \left( \frac{T}{d}
  \right)^{d-1} \sim T^{d-1}\, .
\end{align}
Here $\gamma$ is the metric induced on the horizon from
\eqref{back-geo}. This scaling of the entropy with temperature is an important result to keep in mind
as our primary objective in this work is to determine the
modification of this scaling relation due to the disorder.

\subsection{Scalar solutions at ${\cal O}(\bar V)$}

We now turn on the disorder with strength $\bar V$. That is, we introduce bulk
scalars whose near-boundary behavior, modulo a factor of $r^{d-\Delta}$,
gives the boundary source \eqref{bdy-source}. This scalar solution is
determined by the wave equaton in the background \eqref{back-geo}:
\begin{align}
  0 ={}& r^{d+1} \partial_r \left( r^{-(d-1)} f \partial_r \Phi^{(1)}
  \right) + r^2 \partial_i^2 \Phi^{(1)} + \mu \Phi^{(1)}\, .
\end{align}
Since we are using a spectral representation of the source on the
boundary, we decompose our bulk scalar into harmonics as well,
\begin{align}
  \Phi^{(1)}(r,x^i) ={}& C \bar V \sum_{\{n_i\}} \phi_{k}(r)
  \prod_i \cos (k_{i,n_i} x^i + \gamma_i)\, , \label{eq:phi-spec}
\end{align}
where the $\phi_{k}$ (with $k = |\vec{k}_{n_i}|$) now solve the ODE:
\begin{align}
  0 ={}& r^{d+1} \partial_r \left( r^{-(d-1)} f \partial_r
    \phi_{k} \right) - (k^2 r^2 - \mu) \phi_{k}\, . \label{eq:phik-ODE}
\end{align}
The holographic prescription is to
find the linear combination of solutions to this differential equation
which behave as in \eqref{bdyscalar} near the boundary and are regular
at the horizon $r_+$. That is, due to the various constants that have been
factored out, we are to pick the solutions of \eqref{phik-ODE} that
are regular at the horizon and behave at small $r$ as $\phi_k(r \to 0)
= r^{d-\Delta} + \dotsb$.

This differential equation does not have a closed form
solution for $d>2$, as the emblackening factor introduces $d$ singular
points.\footnote{In the appendix we have the closed form solution for
  $d=2$.} Fortunately, for reasons that will become clear below, we
only need the large $k$ behavior of the scalars. These large $k$ modes will be
responsible for the leading IR singular behavior after disorder averaging. In the large $k$
regime we can employ a WKB approximation to find (letting $\kappa = k r_+$
and $\rho = r/r_+$):
\begin{align}
  \phi_\kappa(\rho) ={}& \frac{\rho^{\frac{d-1}{2}}}{f^{1/4}(\rho)}
  \exp \left[ - \kappa \rho \, {}_2 F_1\left(\tfrac{1}{2},
      \tfrac{1}{d}, 1 + \tfrac{1}{d}, \rho^d \right) \right]\,
  . \label{eq:phiwkb}
\end{align}
The WKB limit here is $\kappa = k r_+ \to \infty$, or $k/T \to
\infty$. These modes are largely insensitive to the presence of the horizon,
decaying well before reaching the horizon, whereas the small $\kappa
\ll 1$ modes will only weakly vary between the boundary and the
horizon.

\subsection{Geometry at ${\cal O}(\bar V^2)$}

Once the scalars are turned on in the bulk, they source the Einstein
equations at order $\bar V^2$. As mentioned above, to start with we
will find the geometry induced by the averaged stress tensor. To leading order
this is the disorder-averaged finite temperature metric.
To that end, we calculate the averaged trace-reversed stress tensor:
\begin{align}
  \kappa_N^2 \Theta_{AB} ={}& 2\Bar{\partial_A \Phi^{(1)} \partial_B
    \Phi^{(1)}} + \frac{4}{d-1} g_{AB} \Bar{V(\Phi)} =
  2\Bar{\partial_A \Phi^{(1)} \partial_B \Phi^{(1)}} -
  \frac{2\mu}{d-1} g_{AB} \Bar{(\Phi^{(1)})^2}\, .
\end{align}
The needed averages are simple to calculate using the spectral
decomposition \eqref{phi-spec}, and the resulting sources are:
\bea
  \kappa_N^2 f^{-1} \Theta_{tt} & = & \frac{\mu \bar V^2 C^2}{2^{d-2} (d-1)}
  \sum_{\{n_i\}} \frac{\phi_k^2}{r^2}\, , \\
  \kappa_N^2 f
  \Theta_{rr} & = & \frac{\bar V^2 C^2}{2^{d-2}} \sum_{\{n_i\}} \left[ f
    (\phi_k')^2 - \frac{\mu}{d-1} \frac{\phi_k^2}{r^2} \right]\,
  , \label{eq:theta-1}  \\
  \kappa_N^2 \Theta_{ii} & = & \frac{\bar V^2 C^2}{2^{d-2}} \sum_{\{n_i\}}
  \left(r^2 k_{i,n_i}^2 - \frac{\mu}{d-1} \right) \frac{\phi_k^2}{r^2}\,
  . \label{eq:theta-2}
\eea
Since we have taken an isotropic distribution, the scalar
sources in the spatial direction $\Theta_{ii}$ are equal for all
$i$. This will simplify the resulting geometry considerably.

With these sources, we search for a perturbative solution for $\bar V
\ll 1$ of the form:
\begin{align}
  \dd s^2 ={}& \frac{1}{r^2} \left[ - f(r) \left(1 + \bar V^2 A(r)
    \right)\dd t^2 + \frac{\dd r^2}{f(r)} + \left(1 + \bar V^2 B(r)
    \right)\sum_{i=1}^{d-1} (\dd x^i)^2 \right]\,
  . \label{eq:metric-ansatz}
\end{align}
Plugging this ansatz into Einstein's equations then yields the
following system of coupled ODEs:
\begin{align}
  0 ={}& f^2 A'' + \frac{df(f-3)}{2r} A' + \frac{(d-1)f(rf'-2f)}{2r}
  B' - 2 \kappa_N^2 \Theta_{tt}\,, \label{eq:efe-tt}\\
  0 ={}& A'' - \left(\frac{1}{r} - \frac{3 f'}{2f}\right) A' + (d-1)
  \left[B'' - \left( \frac{1}{r} - \frac{f'}{2f} \right)B' \right] +
  2\kappa_N^2 \Theta_{rr}\, , \label{eq:efe-rr} \\
  0 ={}& f B'' - \frac{d+ f(d-2)}{r} B' - \frac{f}{r} A' + 2 \kappa_N^2
  \Theta_{ii}\, . \label{eq:efe-ii}
\end{align}
These equations can be re-expressed as a first order constraint
equation plus two decoupled second order equations. The second order
equations are:
\begin{align}
  0 ={}& \frac{2 \left[d+(d-2)f \right]^2}{r^{1-d} f^{1/2}} \partial_r
  \left[ \frac{r^{1-d} f^{3/2}}{d+(d-2)f} \partial_r A \right] +
  j_A(r)\, , \label{eq:aeq} \\
  0 ={}& 2 r^{d-1} f^{1/2} \partial_r \left[
    \frac{f^{1/2}}{r^{d-1}} \partial_r B \right] + j_B(r)\,
  , \label{eq:beq}
\end{align}
where the scalar sources have been repackaged into:
\begin{align}
  j_A(r) ={}& - 2 \kappa_N^2 \left\{ [d+ 3(d-2) f] f^{-1} \Theta_{tt}
    - [d-(d-2)f] \left( f \Theta_{rr} - \sum_i \Theta_{ii} \right)
  \right\}\, ,\label{eq:jadef} \\
  j_B(r) ={}& \frac{2 \kappa_N^2}{d-1} \left[ f^{-1} \Theta_{tt} + f
    \Theta_{rr} + \sum_i \Theta_{ii} \right] \, .\label{eq:jbdef}
\end{align}

The decoupled second order equations (\ref{eq:aeq}) and (\ref{eq:beq})
can be solved exactly. It is useful to work with the rescaled coordinate
\be
\rho \equiv \frac{r}{r_+} \,.
\ee
If we further rescale the sources $r_+^2 j_{A/B} \to j_{A/B}$, then we can write
the solution as:
\begin{align}
  A(\rho) ={}& \alpha_1 + \alpha_2 f^{-1/2} [(d-2) f - d] \nonumber
  \\
  &\quad \quad + \frac{1}{d} \int_\rho^1 \frac{\dd \tilde
    \rho}{\tilde \rho^{d-1}} \left[ \frac{d-(d-2)f(\tilde
      \rho)}{f^{1/2}(\tilde \rho)} - \frac{d-(d-2)
      f(\rho)}{f^{1/2}(\rho)} \right] \frac{ f^{1/2}(\tilde \rho)
    j_A(\tilde \rho)}{[d+(d-2)f(\tilde \rho)]^2}
  \, ,\label{eq:asol}  \\
  B(\rho) ={}& \beta_1 + \beta_2 r_+^d \left[ f^{1/2}(\rho) - 1
  \right] + \frac{1}{d} \int_\rho^1 \frac{\dd \tilde \rho}{\tilde
    \rho^{d-1}} \left[ 1 - \frac{f^{1/2}(\rho)}{f^{1/2}(\tilde \rho)}
  \right] j_B(\tilde \rho)\, . \label{eq:bsol}
\end{align}
Thus we have reduced the scalar backreaction to two  integrals.
There are a number of integration constants in \eqref{asol} and \eqref{bsol}. Indeed, the behavior of
the metric at the horizon is entirely determined by these
constants. Plugging these solutions into \eqref{efe-rr} (or, equivalently, into the first
order equation that can be derived from the three equations above) shows that
$\beta_2$ is fixed by $\alpha_2$, $\beta_2 \propto \alpha_2$ (we only need to say
they are proportional, as they will both be zero shortly). In showing that the remaining
equation is satisfied, it is important to verify that the disorder averaged energy
momentum tensor is conserved.

The constants are fixed by the boundary conditions we impose on the
metric, both at the conformal boundary and at the horizon. The
physical requirements for the geometry are that it is asymptotically
AdS$_{d+1}$ and that it is regular at the horizon. Regularity at the
horizon is easily seen to require $\alpha_2=0$ (and hence
$\beta_2=0$). At the conformal boundary, we require $A(0) = B(0)$. The
actual value of $A(0)$ can be scaled away by redefining coordinates so
we will impose the simple condition $A(0) = B(0) = 0$, that is:
\begin{align}
  \alpha_1 ={}& - \frac{1}{d} \int_0^1 \frac{\dd \rho}{\rho^{d-1}}
  \left[ \frac{d-(d-2)f(\rho)}{f^{1/2}(\rho)} - 2 \right] \frac{
    f^{1/2}(\rho) j_A(\rho)}{[d+(d-2)f( \rho)]^2} \,
  , \label{eq:alpha-sol}\\
  \beta_1 ={}& - \frac{1}{d} \int_0^1 \frac{\dd \rho}{\rho^{d-1}}
  \left[ 1 - \frac{1}{f^{1/2}(\rho)} \right] j_B(\rho)\,
  . \label{eq:beta-sol}
\end{align}
Thus we obtain an explicit expression for the metric at order $\bar V^2$.

\subsection{Large momentum backreaction}

In this section, we will look at the backreaction induced by the large momenta
(relative to the temperature) scalar modes, where the WKB solutions
\eqref{phiwkb} are valid. In particular, we are interested in the
behavior of the metric at the horizon, as this is what determines the
entropy. To get the metric at the horizon, we must determine the integration constants
$\alpha_1$ and $\beta_1$ (because the integrals in (\ref{eq:asol}) and
(\ref{eq:bsol}) vanish at the horizon $\rho = 1$). We will see that these large
momentum modes give the leading singular contribution to $\alpha_1$ and $\beta_1$.

Since the average stress tensor is a decoupled sum of contributions
from each wavevector, we write $j_{A/B} = \sum_{\vec{\kappa}}
j_{A/B,\kappa}$. The large momentum contribution to $\alpha_1,
\beta_1$ can be readily evaluated in the limit $\kappa \to \infty$,
using the WKB solutions \eqref{phiwkb}:
\begin{align}
  \alpha_{1, \kappa} ={}& - \frac{1}{d} \int_0^1 \frac{\dd
    \rho}{\rho^{d-1}} \left[ \frac{d-(d-2)f(\rho)}{f^{1/2}(\rho)} - 2
  \right] \frac{ \rho^{1-d} f^{1/2}(\rho) j_{A,\kappa}(\rho) }{
    [d+(d-2)f( \rho)]^2} \nonumber \\
  ={}& -
  \frac{\kappa^{1-d} \Delta \kappa \Gamma(d+1)}{2^d d(d-1)}\, + {\cal O}(\kappa^{1-2d}) \, , \\
  \beta_{1,\kappa} ={}& - \frac{1}{d} \int_0^1 \frac{\dd
    \rho}{\rho^{d-1}} \left[ 1 - \frac{1}{f^{1/2}(\rho)} \right]
  j_{B,\kappa}(\rho) = \ocal \left(\kappa^{-d} \right)\,
  . \label{eq:largekbeta}
\end{align}
The integrals are performed by noting that at large $\kappa$, the small
$\rho$ (near boundary) region dominates (specifically, $\rho \sim 1/\kappa$).
This means that the WKB solution
(\ref{eq:phiwkb})  goes like $e^{- \kappa \rho}$, while the remaining terms in the
integrand can be expanded about $\rho = 0$.

Recall that $\kappa \equiv k r_+$, so that the WKB limit is $k/T \to \infty$.
We can sum the above large $\kappa$ terms up to find:
\begin{align}
  \sum_{\kappa \gg 1}^{\kappa_0} \alpha_{1,\kappa} & \simeq{} -
  \int_{\kappa \gg 1}^{\kappa_0} \dd^{d-1} \kappa \,
  \frac{\kappa^{1-d} \Gamma(d+1)}{2^d d(d-1)} = - \int^{\kappa_0} \dd
  \kappa\, \frac{\pi^{\frac{d}{2}-1} \Gamma\left( \tfrac{d}{2}
    \right)}{2 \kappa} \nonumber \\ &\simeq - \pi^{\frac{d}{2}-1}
  \Gamma\left( \tfrac{d}{2} \right) \log \kappa_0\, . \label{eq:log}
\end{align}
 In the last step
we have picked out the singular contribution due to the upper endpoint
$\kappa_0 \gg 1$ of the integral.
The large $\k$ contributions to $\beta_1$ in (\ref{eq:largekbeta})
are smaller by a power of $\k$ than the large $\k$ contributions to $\alpha_1$. There is no singular
contribution in that case.

Since $\alpha_1$ and $\beta_1$ will also generically receive non-zero
contributions from all momenta, we write:
\begin{align}
  \alpha_1 ={}& \eta_1 - \pi^{\frac{d}{2}-1} \Gamma \left(
    \tfrac{d}{2} \right) \log \kappa_0\, , & \beta_1 ={}& \eta_2\, .
\end{align}
If $\eta_1$, $\eta_2$ have no singular dependence on $r_+$
in the range $\kappa_0/N \ll 1 \ll \kappa_0$, then this
information is all we need to determine the low temperature scaling of the entropy
density. But this last statement is indeed true. It may be verified 
by numerically evaluating all the necessary integrals. Alternatively, we
can note physically that the only place that such dependence could arise is from the IR
cutoff $\kappa_0/N$; however the modes near the IR cutoff, in the
limit $N \to \infty$, are essentially constant between the boundary and the
horizon and so their contribution will be that of a $\kappa=0$ mode,
which won't introduce any singular $r_+$ dependence.

The logarithmic divergence in (\ref{eq:log}) has essentially the same
origin as the zero temperature logarithm found in \cite{Hartnoll:2014cua}, as well as the
logarithms arising in earlier works \cite{Adams:2011rj, Adams:2012yi}.

\section{Thermodynamics}
\label{sec:therm}

We can easily obtain the thermodynamic properties of the averaged geometries
constructed in the previous section. First we recall that the
temperature of the horizon is determined by the surface gravity, $\hat \kappa$, which
now receives contributions from $\alpha_1$:
\begin{align}
  T ={}& \frac{\hat \kappa}{2\pi} = \frac{d}{4 \pi r_+} + \frac{1}{2}
  f'(r_+) A(r_+) \bar V^2 + {\cal O}(\bar V^4) \\
  ={}& \frac{d}{4 \pi r_+} \left[ 1 + \frac{1}{2} \eta_1 - \frac{1}{2}
    \pi^{\frac{d}{2}-1} \Gamma \left( \tfrac{d}{2} \right) \bar V^2
    \log (k_0 r_+) \right] + {\cal O}(\bar V^4)\, .
\end{align}
If, in the spirit of \cite{Hartnoll:2014cua}, we throw caution to the wind and exponentiate the logarithm, then to
this order we can write:
\begin{align}
  T \sim r_+^{- z}\, ,
\end{align}
where
\begin{align}  
z ={}& 1 +
  \frac{1}{2} \pi^{\frac{d}{2}-1} \Gamma \left( \tfrac{d}{2} \right)
  \bar V^2 + {\cal O}(\bar V^4)\, .\label{eq:zfinal}
\end{align}
This is precisely the dynamical critical exponent identified at $T =0$ in
\cite{Hartnoll:2014cua}. The constant $\eta_1$ has gone into the prefactor
in this scaling relation. 

Now that we know how the temperature scales with the horizon, we can
determine the entropy density scaling:
\begin{align}
  s ={}& \frac{1}{4 G_N} \frac{1}{V} \int\limits_{r=r_+} \dd^{d-1} x \, \sqrt{\gamma} =
  \frac{1}{4 G_N} \frac{r_+^{- (d -1)}}{V} \int \dd^{d-1} x \left[ 1 +
    \frac{d-1}{2} \bar V^2 B(r_+) + {\cal O}(\bar V^4) \right] \, .
\end{align}
Since $B(r_+)$ is simply an $r_+$-independent constant at low temperatures, we see that
this entropy scales as:
\begin{align}
  s \sim r_+^{-(d-1)} \sim T^{\frac{d-1}{z}}\, . \label{eq:s1}
\end{align}
This scaling relation is the first incarnation of our primary result. In particular,
we see that the exponent $z$ is a true critical exponent and that the
disorder has indeed affected thermodynamic properties.
The result (\ref{eq:s1}) relates the entropy and temperature of the averaged metric.
In the following subsection we show that this relation also holds for the true
entropy as a function of temperature.

\subsection{Configuration dependence} 
\label{sec:config}

In this section, we discuss the entropy of a typical, configuration dependent metric. Without the
enhanced symmetry of the averaged configuration to simplify matters, the line element will in
general look like:
\begin{align}
  \dd s^2 ={}& \frac{1}{r^2} \left[- f(r) \left( 1 + \bar V^2 A(x^i,
      r) \right) \dd t^2 + \frac{\dd r^2}{f(r)} + \sum_{ij} \left(1 +
      \bar V^2 B_{ij}(x^i, r) \right) \dd x^i \dd x^j \right]\, .
\end{align}
We cannot solve analytically for the metric functions $A, B_{ij}$ in general.
However, the entropy depends only on the induced metric on
the horizon. To leading order:
\begin{align}
  s ={}& \frac{1}{4 G_N} \frac{1}{V} \int\limits_{r=r_+} \dd^{d-1} x
  \sqrt{\gamma} = \frac{1}{4 G_N} \frac{r_+^{-(d-1)}}{V} \int
  \dd^{d-1} x\sqrt{ \det (1 +
    \bar V^2 B_{ij}(r_+))} \nonumber  \\
  ={}& \frac{1}{4 G_N} \frac{r_+^{-(d-1)}}{V} \int \dd^{d-1} x \left[
    1 + \frac{1}{2} \bar V^2 \sum_i B_{ii}(r_+) \right] = \frac{1}{4
    G_N} \frac{1}{ r_+^{d-1}} \left[ 1 + \frac{1}{2} \bar V^2 \sum_i
    \overline{B_{ii}}(r_+) \right]\, .
\end{align}
Here we have used $\det (1 + \epsilon A) = 1 + \epsilon \tr A + {\cal
  O}(\epsilon^2)$ and the fact that the metric components, being given
by linear combinations of sines and cosines, are self averaging. This
result tells us that, to the order at which we are working, the
entropy density only depends on the averaged spatial metric, and we
can use the result of the previous section to conclude that $s \sim r_+^{-
  (d-1)}$, just as before. Therefore to determine if the entropy of
the averaged metric is distinct from the entropy of a typical configuration
dependent metric all we need to do is find the surface gravity.

The surface gravity is easily worked out to
lowest order in full generality:
\begin{align}
  \hat \kappa^2  = & {} \frac{[f'(r_+)]^2}{4} \left[1 + \bar V^2 A(r_+,x^i) +
    {\cal O}(\bar V^4) \right] \nonumber \\ & \Rightarrow \hat \kappa = \frac{\lvert
    f'(r_+) \rvert}{2} \left[ 1 + \frac{1}{2} \bar V^2 A(r_+,x^i) +
    {\cal O}( \bar V^4) \right]\, .
\end{align}
It is a theorem (for metrics of the form we are considering) that the surface gravity must be constant along the
horizon. Therefore we can replace $A(r_+,x^i)$ in the previous equation by its average:
\begin{align}
  \hat \kappa ={}& \frac{\lvert f'(r_+) \rvert}{2} \left[ 1 + \bar V^2
    \overline{A(r_+)} + {\cal O}(\bar V^4) \right]\, .
\end{align}
More explicitly, averaging $\hat \kappa$ over the
horizon is trivial since it is a constant, whereas averaging
$A(r_+,x^i)$ over space is the same as averaging over the
disorder ensemble. Using our results for the average
metric to we can deduce:
\begin{align}
  \hat \kappa \sim{}& T \sim \frac{\lvert f'(r_+) \rvert}{2} \left[ 1 -
    \frac{1}{2} \pi^{\frac{d}{2} - 1} \Gamma\left( \tfrac{d}{2}
    \right) \bar V^2 \left(\log k_0 r_+ + \text{const.}  \right)
  \right] \sim r_+^{-z} \, ,
\end{align}
where $z$ is again the exponent identified above and in \cite{Hartnoll:2014cua}.

The temperature and entropy scalings, $r_+ \sim T^{-1/z}$ and $s \sim r_+^{-
  (d-1)}$, combine to give:
\begin{align}\label{eq:sfinal}
  s \sim r_+^{-(d-1)} \sim T^{(d-1)/z}\, ,
\end{align}
now as a result for the actual entropy as a function of temperature. In the remainder of the paper
we will verify this result with full blown numerics, beyond the perturbative regime.

\section{Numerics}
\label{sec:num}

In order to construct the fully backreacted black hole solution, at finite disorder strength $\bar V$, we use the DeTurck trick \cite{Headrick:2009pv,Figueras:2011va}. The method works in a general number of dimensions, and we shall use it for $d=2,\,3$ (boundary spacetime dimensions). We will detail the $d=3$ construction, since it is more involved and has not yet appeared in the literature.\footnote{The DeTurk method was used to construct disordered spacetimes in \cite{Hartnoll:2014cua} and \cite{Donos:2014yya}. Other numerical studies of strong disorder in holography have been in the probe limit \cite{Arean:2013mta, Zeng:2013yoa, Arean:2014oaa}.}

The black hole solution we search for is static, which means we can introduce an adapted coordinate system in which $\partial_t$ is a Killing direction. Furthermore, the line element should be invariant under the discrete transformation $t\to-t$. The most general
line element and scalar field compatible with these symmetries can be written
\begin{multline}
\mathrm{d}s^2 = \frac{1}{y^2}\Bigg[-(1-y^3)A\,y_+^2\,\mathrm{d}t^2+\frac{B}{1-y^3}\mathrm{d}y^2+\\
y_+^2 S_1\Big(\mathrm{d}x_1+F_1 \mathrm{d}y+F_2 \mathrm{d}x_2\Big)^2+y_+^2\,S_2\Big(\mathrm{d}x_2+F_3 \mathrm{d}y\Big)^2\Bigg]\,,
\label{eq:metric3D}
\end{multline}
\begin{equation}
\Phi =\frac{y}{y_+}\,\widetilde{\Phi}\,,
\label{eq:phie}
\end{equation}
where $A,\,B,\,S_1,\,S_2,\,F_1,\,F_2,\,F_3$ and $\widetilde{\Phi}$ comprise a total of $8$ functions that depend on $y$, $x_1$ and $x_2$. The first step in using the DeTurck method is to choose a reference metric $\bar{g}$. This reference metric should have our desired boundary conditions (\emph{i.e.} contains a regular horizon and has the correct asymptotics). For the reference metric, we choose the planar Schwarzschild black hole, which can be obtained from the line element (\ref{eq:metric3D}) by setting $A=B=S_1=S_2=1$ and $F_1=F_2=F_3=0$\,. The reference metric does not depend on the boundary spatial coordinates $x_1$ and $x_2$, and so is automatically periodic with respect to these. Finally, $y_+$ is a parameter which controls the black hole temperature: $4\pi T= 3\,y_+$.

The second step in the DeTurck method consists of solving the following set of equations
\begin{align}
 G_{AB}\equiv R_{AB} -\nabla_{(A}\xi_{B)}-& 2 \nabla_A \Phi
  \nabla_B \Phi -\frac{1}{d-1} g_{AB} \left[ 4V(\Phi)+\Lambda\right]=0\, . \nonumber \\
  \Box \Phi - V'(\Phi) = 0\, ,
  \label{eq:dennis}
\end{align}
where  $\xi^M = g^{PQ}[\Gamma^M_{PQ}(g)-\bar{\Gamma}^M_{PQ}(\bar{g})]$ and $\bar{\Gamma}(\bar{g})$ is the Levi-Civita connection associated with the reference metric $\bar{g}$. Furthermore, $V(\Phi)$ is given by (\ref{eq:vphi}) with a mass saturating the Harris criterion (\ref{eq:harris}).
The advantage of this method is that the above Eqs.~(\ref{eq:dennis}) form a set of elliptic PDEs \cite{Headrick:2009pv} for the metric ansatz (\ref{eq:metric3D}), unlike the Eqs.~(\ref{eq:einstein}).  For solutions of Eq.~(\ref{eq:dennis}) to also be solutions of Eq.~(\ref{eq:einstein}), we must have $\xi^M=0$.  In some cases (such as vacuum Einstein), there is a proof that all solutions to Eq.~(\ref{eq:dennis}) also have $\xi^M=0$ \cite{Figueras:2011va}.  In our case, we lack such a proof, so we must verify that $\xi^M=0$ after solving the equations. The local uniqueness property of elliptic equations guarantees that solutions with $\xi^M\neq0$ cannot be arbitrarily close to those with $\xi^M=0$.

In order to complete the system of partial differential equations, suitable boundary conditions must be imposed. In addition, these must be consistent with zero DeTurck vector $\xi^M$. At the boundary, located at $y=0$, we demand $A(0,x_1,x_2)=B(0,x_1,x_2)=S_1(0,x_1,x_2)=S_2(0,x_1,x_2)=1$ and $F_1(0,x_1,x_2)=F_2(0,x_1,x_2)=F_3(0,x_1,x_2)=0$. Furthermore, we demand $\tilde{\Phi}(0,x_1,x_2)=\Phi_1(x_1,x_2)$, with the scalar source $\Phi_1(x_1,x_2)$ defined in Eq.~(\ref{eq:bdy-source}). The reader might be surprised with the extra factor in Eq.~(\ref{eq:phie}) dependent on $y_+$. However, we note that, asymptotically, the relation between the Fefferman-Graham coordinate $r$ defined in Eq.~(\ref{eq:fgcoordinates}) and $y$ reads $y = y_+\, r+\mathcal{O}(r^2)$.

At the horizon, $y=1$, the Einstein-DeTurck equations demand $A(1,x_1,x_2)=B(1,x_1,x_2)$, which is equivalent to having a well defined bifurcating Killing horizon, with our choice of reference metric. The boundary conditions at the horizon for the remaining variables follow from expanding the equations in a power series off the horizon - they all turn out to be of the Robin type. For the $x_1$ and $x_2$ directions, we demand periodic boundary conditions.

Now we are in a position to solve the PDE system (\ref{eq:dennis}) subject to the above mentioned boundary conditions. To solve the equations, we use a standard damped Newton-Raphson iteration algorithm based on pseudo-spectral collocation on a Chebyshev (in $y$) and Fourier grids (in $x_1$ and $x_2$). In $d=2$, there are additional subtleties associated with the boundary behaviour of the scalar field $\Phi$, which introduces several non-analytic behaviours in the metric and scalar field functions. To deal with these, we patch a grid of finite differences onto the Chebyshev collocation grid parametrising the holographic radial direction $y$ \cite{Hartnoll:2014cua}.

\begin{figure}[h]
\centering
\subfigure{\label{fig:0a} (a) \includegraphics[height = 0.26\textheight]{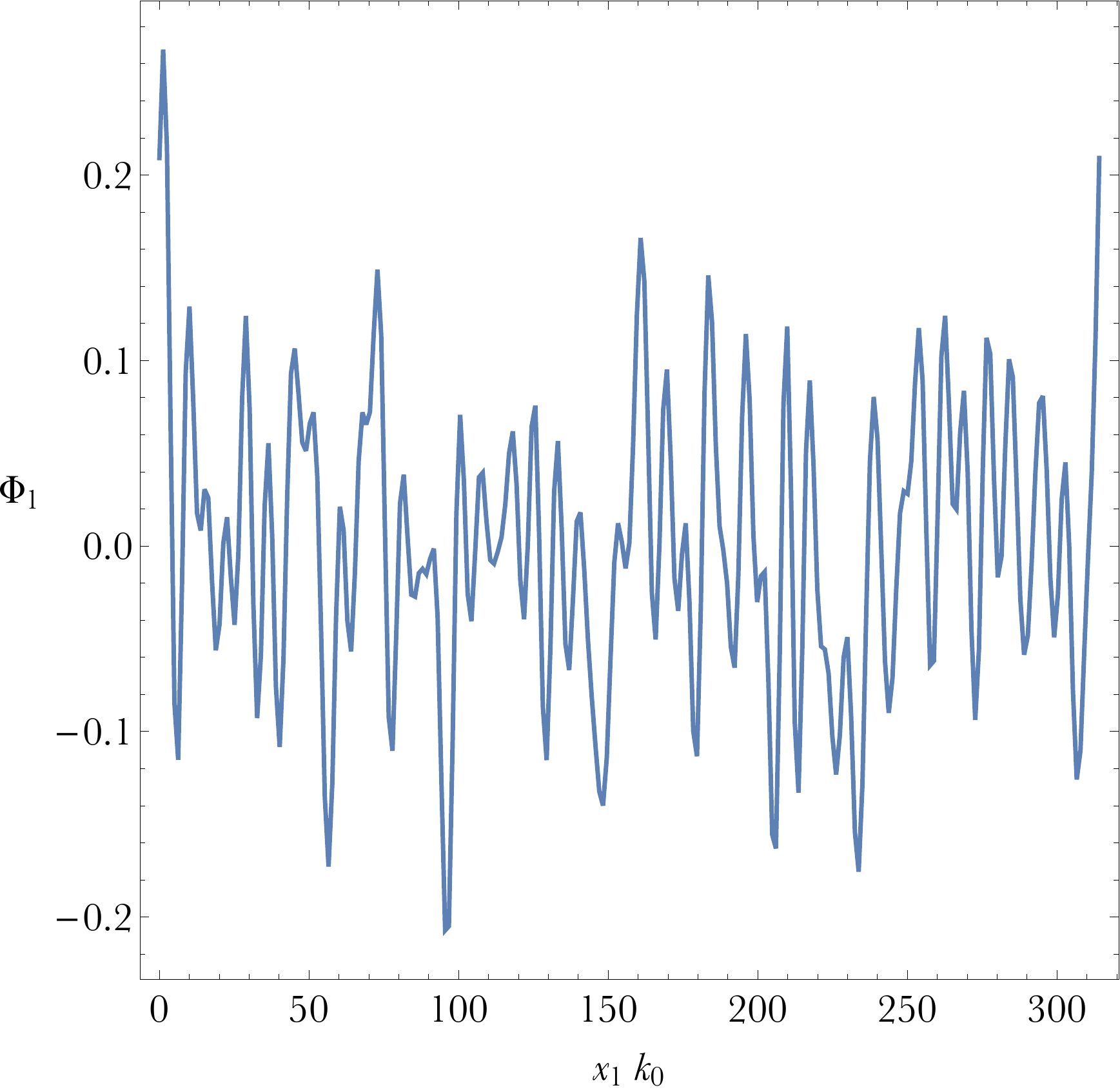}}
\subfigure{\label{fig:0b} (b) \includegraphics[height = 0.26\textheight]{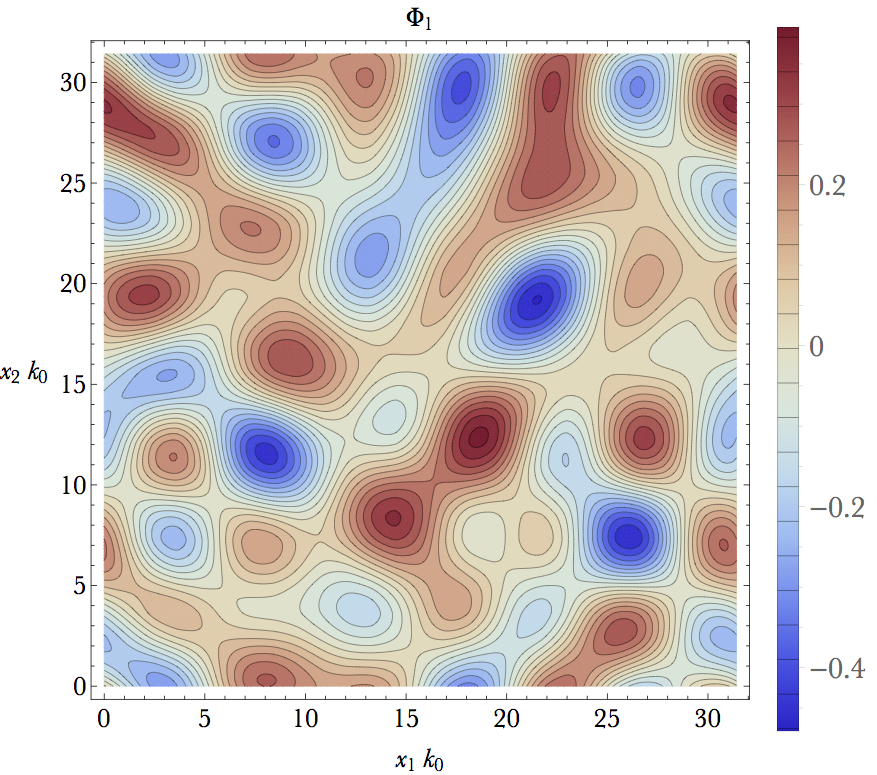}}
\caption{\label{figs:0} {\bf Disordered sources:} Plot (a) shows a scalar source $\Phi_1$ as a function of $x_1\,k_0$ at the boundary. Plot (b) is a density plot of $\Phi_1$, now in $d=3$, as a function of boundary directions $x_1\,k_0$ and $x_2\,k_0$. In both cases we have chosen $\bar{V}=0.1$. The characteristic width of the peaks in these plots is determined by the short distance cutoff, $\Delta x |_\text{peak} \sim \pi/k_0$.}
\end{figure}

The space of solutions is 4-dimensional, depending on $\bar{V}$, $k_0$, $N$ and $T$. However, since our underlying UV microscopic theory is conformally invariant, we only need to worry about dimensionless ratios of these quantities. In order to access the true IR physics, we need to preserve the hierarchy presented in Eq.~(\ref{eq:hierarchy}). That is, we must make sure the temperature range we probe is in between the short and long distance cutoffs on the disorder distribution.
In the Schwarzschild background with $\bar V = 0$, the temperature is given by $T = d/(4 \pi r_+)$, and it is $r_+$ rather than $1/T$ that sets the scale that should be compared to cutoffs. This is helpful because it pushes the IR cutoff down to lower temperatures (by a factor of $d/(4 \pi)$) than the rough window (\ref{eq:hierarchy}) would suggest.\footnote{Having the IR cutoff on the disorder be behind the horizon also resolves the following technical issue that arises at $T=0$. While the disorder is marginally relevant, the homogeneous mode of the scalar is strongly relevant. In the energy range (\ref{eq:hierarchy}) the growth of the homogeneous mode is subdominant to the disorder physics due to the presence of many higher harmonics. However, below the IR cutoff on the disorder, the homogeneous mode will eventually dominate and drive a flow away from AdS. This is an artifact of needing to work with an IR cutoff, and can complicate zero temperature numerics, but not the numerics herein.} In all computations detailed in this section we measure all quantities in units of $k_0$ (which effectively sets $k_0=1$) and either we take $N = 50$ in $d=2$, or $N=5$ in $d=3$. $T$ is then allowed to vary freely in the required range (\ref{eq:hierarchy}). Note that for these values of $N$, the Fourier grids must be very dense, having at least $500$ points in the periodic direction in $d=2$ and $50$ in $d=3$.\footnote{The choice of number of points in each periodic direction is such that we should be able to resolve up to the fifth harmonic of the highest wave number appearing in our scalar field potential (\ref{eq:bdy-source}). Since each harmonic descendent decays exponentially \cite{Horowitz:2012ky} in multiples of the relevant wavenumber, we expect our resolution to capture all the relevant physics.} Typical profiles for the boundary source in $d=2$ and $d=3$ are depicted in Fig.~(\ref{figs:0}).

\begin{figure}[h]
\centering
\subfigure{\label{fig:1a} (a) \includegraphics[height = 0.26\textheight]{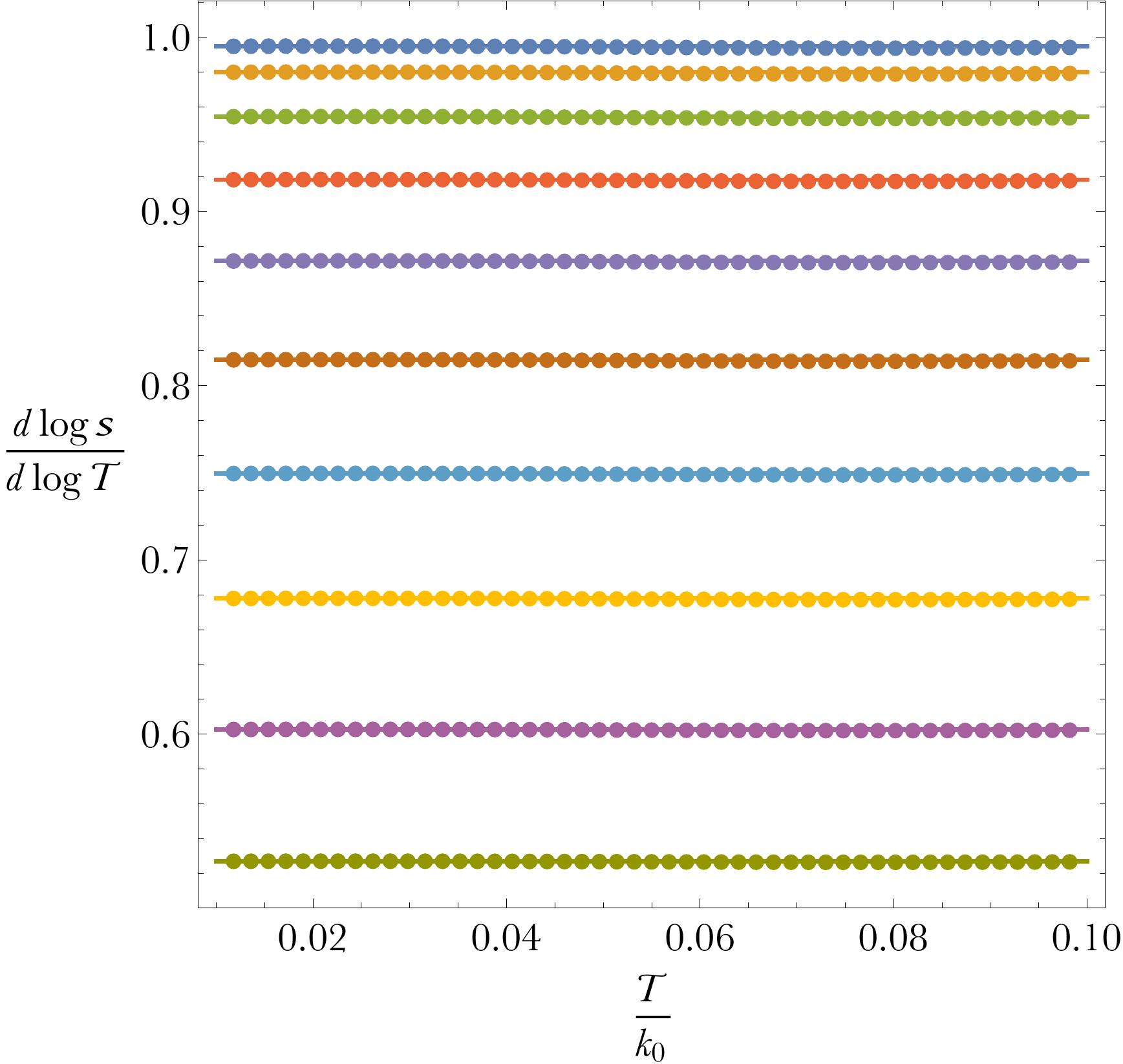}}
\subfigure{\label{fig:1b} (b) \includegraphics[height = 0.26\textheight]{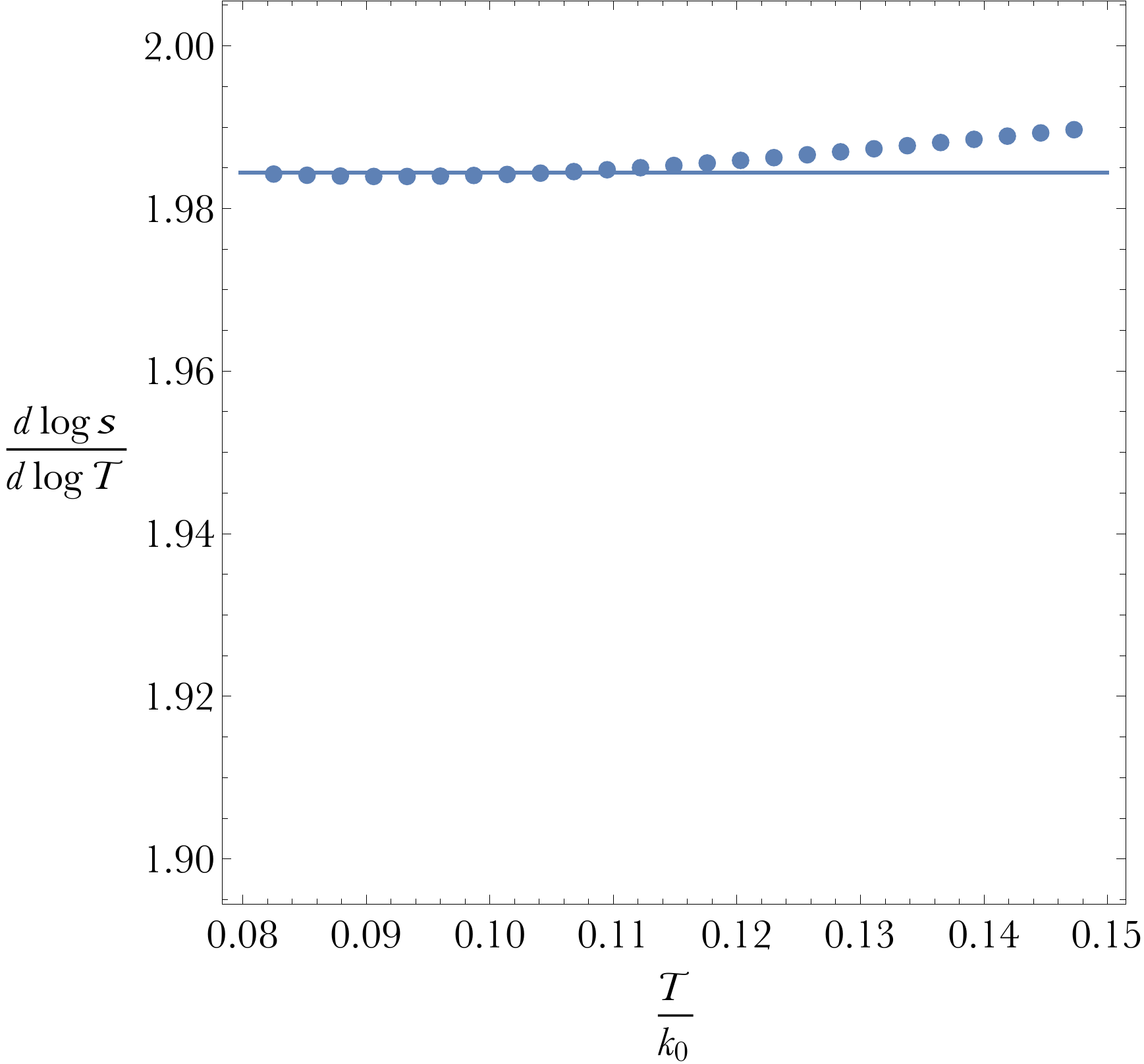}}
\caption{\label{fig:1} {\bf Emergence of an IR dynamical scaling exponent}. Plot (a) shows the logarithmic derivative of the entropy for several values of $\bar{V}$ in $d=2$. These plots have $N=50$. From top to bottom, we have $\bar{V}=0.1,0.2,\ldots,1.0$. Plot (b) shows the logarithmic derivative of the entropy for $\bar{V}=0.1$ in $d=3$. This plot has $N=5$.}
\end{figure}

Having these solutions at hand, there are several quantities we can monitor. We decided to focus on the entropy, since it is a direct probe of the infrared geometry. We will discuss the results for $d=2$ and $d=3$ separately, starting with $d=2$. In Fig.~(\ref{fig:1a}) we show a plot of the logarithmic derivative of the entropy, as a function of the black hole temperature, for several values of the disorder amplitude $\bar{V}$. The data is represented by disks, and the solid lines indicate the analytic prediction of Eq.~(\ref{eq:sfinal}), namely $s \propto T^{(d-1)/z}$ with $z$ given by (\ref{eq:zfinal}). From top to bottom, we have $\bar{V}=0.1,0.2,\ldots,1.0$. The agreement between our perturbative analytic prediction and the numerics is striking. This numerical scaling result is compelling evidence for the emergence of a disordered fixed point at $T=0$, characterized by a dynamical critical exponent $z$.

In $d=3$ the calculations are more involved, since we have to generate many solutions to the 3D PDE system we described above. This means we do not have as much data as for the $d=2$ case. In particular, we have focussed on a single value $\bar{V}=0.1$. We also have a narrower window of temperatures in which to access the IR scaling regime (\ref{eq:hierarchy}) because $N$ is smaller.
 In Fig.~\ref{fig:1b} we plot the logarithmic derivative of the entropy, as a function of the black hole temperature for $\bar{V}=0.1$. The disks represent the data, and the solid line indicates the analytic prediction $s \propto T^{(d-1)/z}$ of (\ref{eq:sfinal}), with $z$ again given by (\ref{eq:zfinal}).
Again, the agreement as the temperature is lowered is rather encouraging.

The computations of the entropy we have discussed are for a given realization of disorder. This is the entropy we have been after. With the numerical data at hand, we can compare this (physical) entropy with the entropy of the averaged metric, as discussed in previous sections. The averaged metric is easily obtained from the numerics by integrating over $x$ and $y$ (see the more extended discussion in \cite{Hartnoll:2014cua}). To compare the averaged entropy with the entropy of the averaged metric, we did the following: we computed the entropy of the average metric and the entropy of the full metric. We subtracted one from the other and found a maximum disagreement around $1\%$. We then did a similar subtraction, but this time for the dynamical critical exponent measured with both entropies, and we found a maximum disagreement of $10^{-4}\%$, which is well within the error in our integration scheme in $d=3$. We take this as strong evidence that the dynamical critical exponent yields the same value whether measured with the entropy of the average or full metric, as we have argued in the previous section. This result also substantiates the claim in \cite{Hartnoll:2014cua} that the averaged metric is a useful bulk quantity to identify the scaling properties of the IR fixed point.

\begin{figure}[h]
\centering
\subfigure{\label{fig:3a} (a) \includegraphics[height = 0.26\textheight]{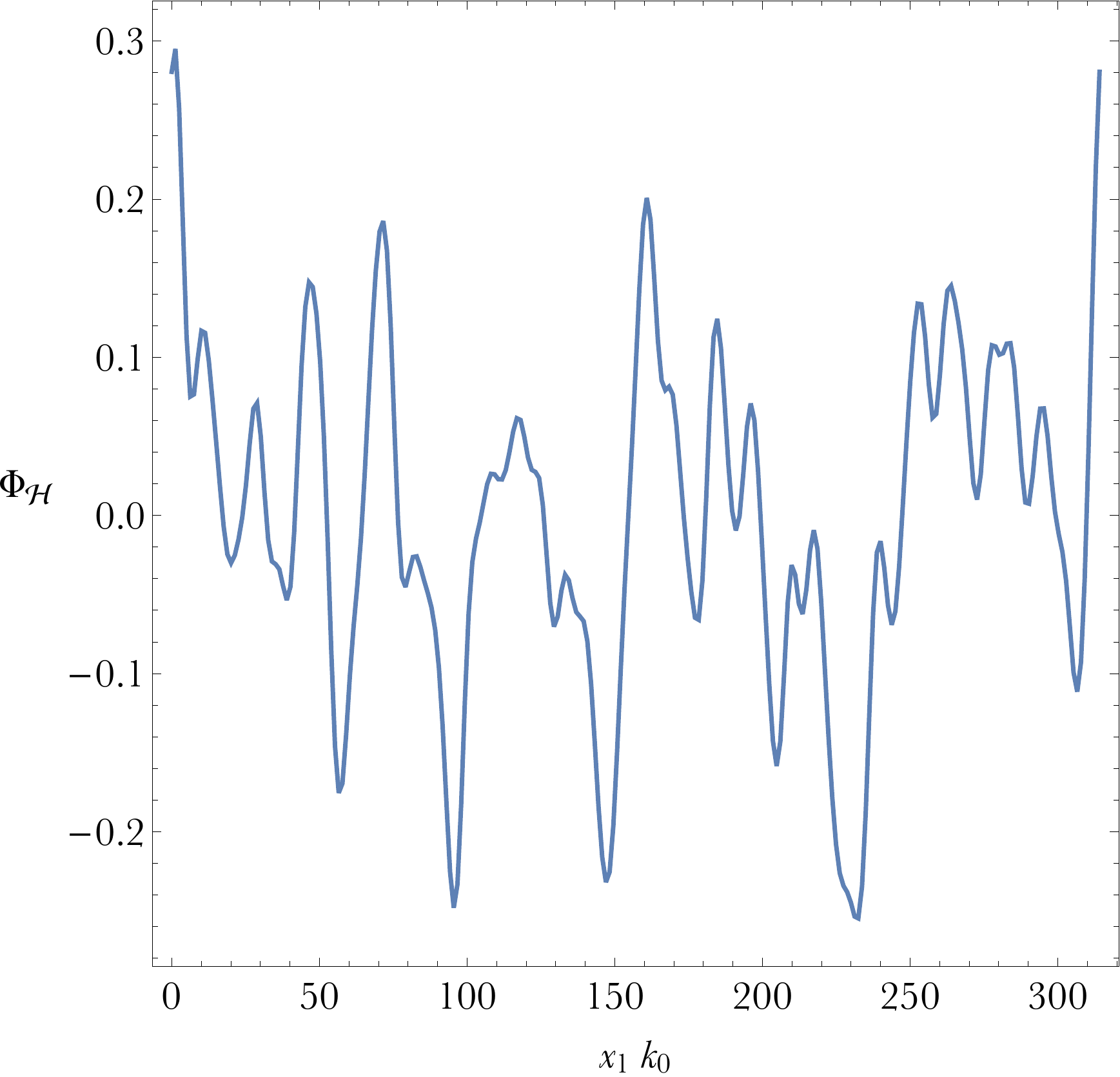}}
\subfigure{\label{fig:3b} (b) \includegraphics[height = 0.26\textheight]{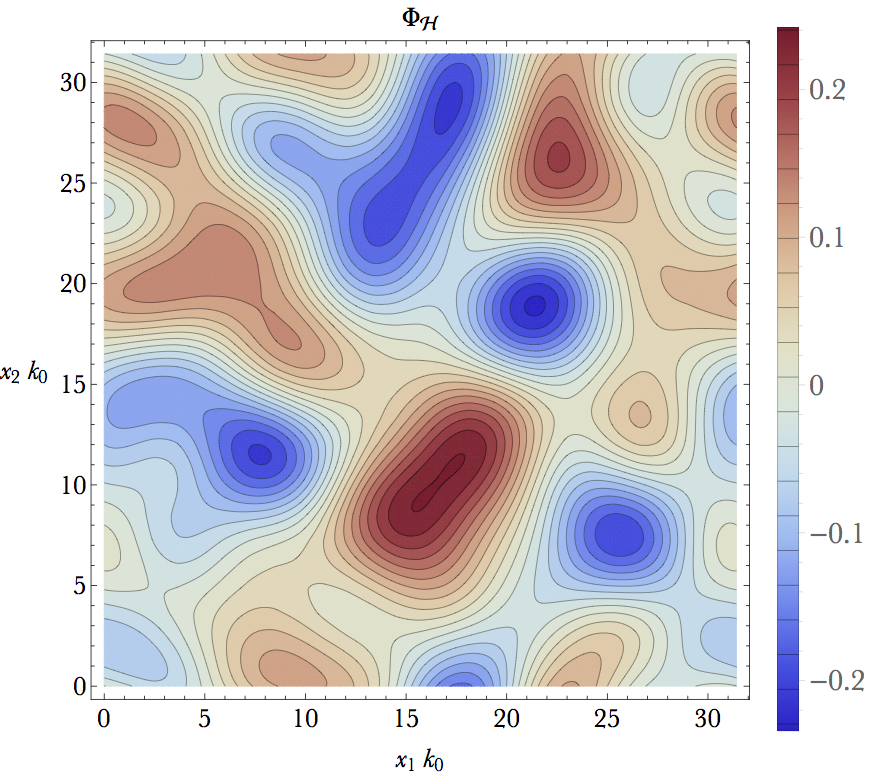}}
\caption{\label{figs:3} {\bf Disordered horizons}. Plot (a) shows the scalar field $\Phi_\mathcal{H}$ evaluated at the horizon as a function of $x_1\,k_0$. Plot (b) is a density plot of $\Phi_\mathcal{H}$, now in $d=3$, as a function of boundary directions $x_1\,k_0$ and $x_2\,k_0$. The sources for these solutions are those shown in Fig.~\ref{figs:0}. Plot (a) is at temperature $T/k_0 = 0.0478$ while plot (b) has $T/k_0=0.0798$. The characteristic width of the peaks in these plots is now determined by the temperature scale, so that $\Delta x |_\text{peak} \sim \pi r_+$. This is the expected statement that the temperature serves as the short distance cutoff on the disorder distribution at the horizon.}
\end{figure}

The reader might also be interested in the difference between the sources shown in Figs.~(\ref{figs:0}) and the scalar field evaluated at the horizon, $\Phi_\mathcal{H}$. For completeness, we show the latter in Figs.~\ref{figs:3}. We see that some of the structure of the UV clearly survives in the IR. At first sight, the disorder appears to have been smoothened out on the horizon relative to the sources shown in Fig.~(\ref{figs:0}). However, this is simply the fact that upon renormalizing down to the horizon,
structure on scales smaller than the temperature scale have been integrated out. This illustrates the need to keep the long distance cutoff on the disorder distribution sufficiently large in order to access the correct disorder physics in the IR geometry.

\begin{figure}[h]
\centering
\includegraphics[height = 0.3\textheight]{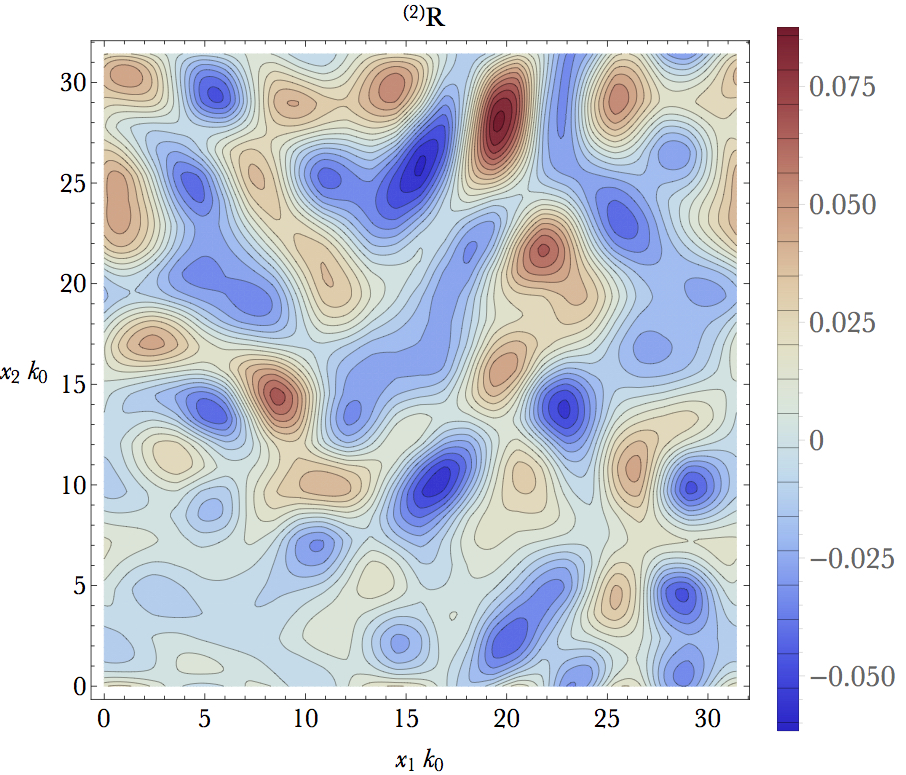}
\caption{\label{fig:4} {\bf Disordered horizon - Ricci scalar}. Plot of the Ricci scalar of the induced metric on a spatial cross section of the horizon. The parameters used are the same as in Fig.~\ref{figs:3}. Because the metric depends on the square of the scalar field, the metric functions oscillate twice as quickly and hence the structures appear half the size of those in Fig.~\ref{figs:3}.}
\end{figure}

Finally, in an attempt to characterize the geometry of the horizon, we plot in Fig.~(\ref{fig:4}) the Ricci scalar, ${}^{(2)}R$, of the induced metric on a spatial cross section of the horizon - {\bf a disordered horizon}.

\section{Discussion}
\label{sec:diss}

In this paper we have presented evidence for the existence of a disordered fixed point in the far IR of a spacetime with a marginally relevant
disordered boundary source. In addition to constructing numerical disordered black hole spacetimes, we showed that the dynamical critical exponent
$z$ of the IR fixed point could be obtained by resumming logarithmic divergences that appear in perturbation theory in the strength of the disorder.
However, this perturbation theory is an expansion about the UV spacetime. The whole point of fixed points is that they are self-contained and well-defined
without reference to a UV completion. Therefore, an intrinsic description of the disordered horizon (at $T=0$) as a solution to Einstein's equations should exist.
Characterizing the disordered horizon on its own, IR, terms could potentially lead to a greatly expanded understanding of what extremal horizons can look like.
It would, presumably, explain why a na\"ive resummation of logarithmic divergences at low orders in perturbation theory appears to give the correct answer
for the dynamical critical exponent. It would also clarify exactly which quantities can be accurately determined from the corresponding disorder averaged spacetime.

Given the construction of the background geometries, it is very natural to study correlation functions in these backgrounds. If the full frequency and momentum
dependence can be found, then this should verify the value of $z$ that we have found, giving correlators that are scaling functions of $\omega/k^z$. The study of
transport, in particular, in these backgrounds was one of the motivations to construct these solutions in the first place. The disordered fixed point does not conserve momentum and will therefore have intrinsically finite transport properties \cite{Hartnoll:2014lpa}. This is a qualitatively distinct regime from the case in which disorder can be described as an irrelevant perturbation about a clean IR fixed point \cite{Hartnoll:2007ih, Hartnoll:2008hs, Hartnoll:2012rj, Lucas:2014sba, Lucas:2015vna}.

Finally, having developed the numerical and analytical methods needed to understand disordered spacetimes, we may soon be in a position to tackle the more difficult case of relevant (rather than marginally relevant) disorder.

\section*{Acknowledgements}

We are grateful to Aristomenis Donos and Veronica Hubeny for helpful comments. SAH is partially supported by a DOE Early Career Award, a Sloan fellowship and the Templeton foundation. This work was undertaken on the COSMOS Shared Memory system at DAMTP, University of Cambridge operated on behalf of the STFC DiRAC HPC Facility. This equipment is funded by BIS National E-infrastructure capital grant ST/J005673/1 and STFC grants ST/H008586/1, ST/K00333X/1.

\appendix

\section{$d=2$ solution}

In $d=2$, one can go (slightly) further in describing the geometry of individual configurations
in terms of some explicit indefinite integrals. The first key difference
is that in $d=2$ the scalar solution at finite temperature can be
written explicitly:
\begin{align}
  \Phi^{(1)}(r,x) ={}& \bar V \sum_{n=1}^{N-1} \phi_n(r) \cos (k_n x +
  \gamma_n)\, , \label{eq:d3-scalar}
\end{align}
where:
\begin{align}
  \phi_n(r) = C\sqrt{\frac{1}{\pi}} \lvert \Gamma \left( \tfrac{3}{4} -
    \tfrac{i k_n r_+}{2} \right) \rvert^2 \sqrt{r} P_{- \frac{1}{2}
    + i k_n r_+} \left( \tfrac{r}{r_+} \right)\, . \label{eq:d3-phin}
\end{align}
Here $P_\nu(x)$ is the Legendre function (for the particular values of
the index $\nu= - \frac{1}{2} + i \kappa$ these are also known as
parabolic cylinder functions). Throughout
this section we will use the notation:
\begin{align}
  \theta_n ={}& k_n x + \gamma_n\, , & \theta_{nm}^\pm ={}& \theta_n
  \pm \theta_m\, .
\end{align}

A further simplification in the $d=2$ case is simply that the spatial
metric is much less complicated when there is only one spatial
direction. The order $\bar V^2$ geometry is given as:
\begin{align}
  \dd s^2 ={}& \frac{1}{r^2} \left[ - f(r) \left( 1 + \bar V^2 A(r,x)
    \right) \dd t^2 + \frac{\dd r^2}{f(r)} + \left( 1 + \bar V^2 B(r,x)
    \right) \dd x^2 \right]\, .
\end{align}
The equations of motion for $A,B$ are particular combinations of
Einstein's equations:
\begin{align}
  0 ={}& 2 f^{1/2} \left[\partial_r \partial_x ( f^{1/2} A ) +
      4 f^{1/2} \partial_r \Phi^{(1)} \partial_x \Phi^{(1)}
    \right]\, , \label{eq:d3-ee1} \\
    0 ={}& r f^{- 1/2} \partial_r \left[
      \frac{f^{3/2}}{r} \partial_r A \right] + \frac{2}{r^2}
    \left[ r^2 f \big( \partial_r \Phi^{(1)} \big)^2 - r^2
      \big( \partial_x \Phi^{(1)} \big)^2 - \mu \big(\Phi^{(1)}\big)^2
    \right] \, , \label{eq:d3-ee2} \\
    0 ={}& r f^{1/2} \partial_r \left[ \frac{f^{1/2}}{r} \partial_r
      B \right] + \frac{2}{r^2} \left[r^2 f \big( \partial_r
      \Phi^{(1)} \big)^2 + r^2 \big( \partial_x \Phi^{(1)} \big)^2 -
      \mu\big(\Phi^{(1)}\big)^2 \right] \,
    , \label{eq:d3-ee3} \\
    0 ={}& r^3 f^{-1/2} \partial_r \left[ \frac{f^{3/2}}{r} \partial_r
      A \right] + r^2 \partial_x^2 A - r f \partial_r
    A - r \partial_r B - 4 \mu (\Phi^{(1)})^2\,
    . \label{eq:d3-ee4}
\end{align}
Plugging in the scalar solution \eqref{d3-scalar} and \eqref{d3-phin},
it is straightforward to determine:
\begin{align}
  A(r,x) ={}& \alpha_1(r) + f^{-1/2}(r) \alpha_2(x) + \tilde
  A(r,x) \, , \label{eq:d3-asol}
\end{align}
where:
\begin{align}
  \tilde A(r,x) ={}& \sum_m a^1_m(r) \cos 2 \theta_m + \sum_{m
    \neq n} a^2_{mn}(r) k_m \left( \frac{\cos \theta_{mn}^+}{k_{mn}^+}
    + \frac{\cos \theta_{mn}^-}{k_{mn}^-} \right)\,
  , \label{eq:A2til-sol}
\end{align}
where the radial profiles for the harmonics are:
\begin{align}
  a^1_m(r) ={}& f^{-1/2}(r) \int_r^{r_+} \dd \tilde r \,
  f^{1/2}(\tilde r) \phi_m'(\tilde r) \phi_m(\tilde r)\, , \\
  a^2_{mn} ={}& 2 f^{-1/2}(r) \int_r^{r_+} \dd \tilde r\,
  f^{1/2}(\tilde r) \phi_m (\tilde r) \phi_n'(\tilde r) \, .
\end{align}
The function $\alpha_1(r)$ in \eqref{d3-asol} is given by integral
expressions similar to what we found for general $d$:
\begin{multline}
  \alpha_1(r) = \eta_1 + \eta_2r_+^2 \left[ f^{-1/2}(r) - 1 \right]
  \\-\sum_m \int_r^{r_+} \frac{\dd \tilde r}{\tilde r^3 } \left[1-
    \sqrt{f(\tilde r)/f(r)} \right] \left[ \tilde r^2 f(\tilde r)
    \left(\phi_m'(\tilde r)\right)^2 - (\tilde r^2 k^2 + \mu) \phi_m^2
  \right]\, . \label{eq:d3-alpha-1}
\end{multline}

Turning now to the spatial metric component, it can be written as some
integration functions and a particular solution:
\begin{align}
  B(r,x) ={}& \beta_1(x) + \beta_2(x) r_+^2 \left[f^{1/2}(r) -1
  \right]+ \tilde B(r,x)\, . \label{eq:d3-B2sol}
\end{align}
The particular solutions can be written as:
\begin{align}
  \tilde B(r,x) ={}& \sum_m \left[b^0_m(r) + b^1_m(r) \cos^2
    \theta_m \right] + \sum_{m\neq n} \left[ b^2_{mn}(r) \cos
    \theta_{mn}^+ + b^3_{mn}(r) \cos \theta_{mn}^- \right]\, ,
\end{align}
where:
\begin{align}
  b^0_m(r) = 2 \int_r^{r_+} \frac{\dd \tilde r\, \left[ f^{1/2}(r) -
      f^{1/2}(\tilde r) \right]}{f^{1/2}(\tilde r)
    \tilde r^3} {}&\left[\tilde r k_m \phi_m(\tilde r)\right]^2\, ,\\
  b^1_m(r) = 2 \int_r^{r_+} \frac{\dd \tilde r\, \left[ f^{1/2}(r)
      - f^{1/2}(\tilde r) \right]}{f^{1/2}(\tilde r) \tilde r^3} {}&
  \left\{f(\tilde r) [\tilde r \phi_m'(\tilde r)]^2 -
    (\tilde r^2 k_m^2 +\mu) \phi_m^2(\tilde r) \right\}\, ,\\
  b^2_{mn}(r) = \int_r^{r_+} \frac{\dd \tilde r\, \left[ f^{1/2}(r)
      - f^{1/2}(\tilde r) \right]}{f^{1/2}(\tilde r) \tilde r^3} {}&
  \left\{\tilde r^2 \left[f(\tilde r)\phi_m'(\tilde r) \phi_n'(\tilde
      r) - k_m k_n \phi_m(\tilde r) \phi_n(\tilde r)\right]
  \right. \nonumber \\
  &\left.\quad -\mu \phi_m(\tilde r) \phi_n(\tilde r) \right\}\, ,
  \\
  b^2_{mn}(r) = \int_r^{r_+} \frac{\dd \tilde r\, \left[ f^{1/2}(r)
      - f^{1/2}(\tilde r) \right]}{f^{1/2}(\tilde r) \tilde r^3} {}&
  \left\{\tilde r^2 \left[f(\tilde r)\phi_m'(\tilde r) \phi_n'(\tilde
      r) + k_m k_n \phi_m(\tilde r) \phi_n(\tilde
      r)\right]\right. \nonumber \\
  &\left. \quad -\mu \phi_m(\tilde r) \phi_n(\tilde r) \right\}\, .
\end{align}

There are a few integration functions left we need to fix. In
particular, the constants $\eta_1, \eta_2$ and the functions
$\alpha_2(x), \beta_1(x)$, $\beta_2(x)$. As we discussed before,
$\eta_1$ receives a logarithmic contribution from the WKB modes, which
works out to be:
\begin{align}
  \eta_1 ={}& - \frac{1}{2} \bar V^2 \log k_0 r_+ + \dotsb\, ,
\end{align}
while $\eta_2, \alpha_2$ and $\beta_2$ are all forced to vanish by
regularity at the horizon. This leaves $\beta_1(x)$, which is
determined by fixing $A(0,x) = B(0,x)$. Our solutions for $A$ and $B$
allow one to write explicit integral expressions for $\beta_1$, though
they do not appear to be illuminating so we have not written them out.


\begin{thebibliography}{99}
 
 \bibitem{harris}
A.~B.~Harris, ``Effect of random defects on the critical behaviour of Ising models,''
 J. of Phys. C {\bf 7}, 1671 (1974).
 
\bibitem{sachdev}
S.~Sachdev, {\it Quantum phase transitions}, CUP 1999.

\bibitem{Hartnoll:2014lpa} 
  S.~A.~Hartnoll,
  ``Theory of universal incoherent metallic transport,''
  Nature Phys.\  {\bf 11}, 54 (2015)
  [arXiv:1405.3651 [cond-mat.str-el]].

\bibitem{Hartnoll:2014cua} 
  S.~A.~Hartnoll and J.~E.~Santos,
  ``Disordered horizons: Holography of randomly disordered fixed points,''
  Phys.\ Rev.\ Lett.\  {\bf 112}, 231601 (2014)
  [arXiv:1402.0872 [hep-th]].
 
\bibitem{Kachru:2008yh} 
  S.~Kachru, X.~Liu and M.~Mulligan,
  ``Gravity duals of Lifshitz-like fixed points,''
  Phys.\ Rev.\ D {\bf 78}, 106005 (2008)
  [arXiv:0808.1725 [hep-th]].
  
\bibitem{Hartnoll:2009sz} 
  S.~A.~Hartnoll,
  ``Lectures on holographic methods for condensed matter physics,''
  Class.\ Quant.\ Grav.\  {\bf 26}, 224002 (2009)
  [arXiv:0903.3246 [hep-th]].

\bibitem{Shinozuka1991}  
  M.~Shinozuka and G.~Deodatis, ``Simulation of stochastic processes by spectral representation,'' Appl. Mech. Rev. {\bf 44} (1991) 191.

\bibitem{Adams:2011rj} 
  A.~Adams and S.~Yaida,
  ``Disordered Holographic Systems I: Functional Renormalization,''
  arXiv:1102.2892 [hep-th].

\bibitem{Adams:2012yi} 
  A.~Adams and S.~Yaida,
  ``Disordered Holographic Systems II: Marginal Relevance of Imperfection,''
  arXiv:1201.6366 [hep-th].
  
\bibitem{Headrick:2009pv} 
  M.~Headrick, S.~Kitchen and T.~Wiseman,
``A New approach to static numerical relativity, and its application to Kaluza-Klein black holes,''
  Class.\ Quant.\ Grav.\  {\bf 27}, 035002 (2010)
  [arXiv:0905.1822 [gr-qc]].
  
\bibitem{Figueras:2011va} 
  P.~Figueras, J.~Lucietti and T.~Wiseman,
``Ricci solitons, Ricci flow, and strongly coupled CFT in the Schwarzschild Unruh or Boulware vacua,''
  Class.\ Quant.\ Grav.\  {\bf 28}, 215018 (2011)
  [arXiv:1104.4489 [hep-th]].
  
\bibitem{Donos:2014yya} 
  A.~Donos and J.~P.~Gauntlett,
  ``The thermoelectric properties of inhomogeneous holographic lattices,''
  JHEP {\bf 1501}, 035 (2015)
  [arXiv:1409.6875 [hep-th]].
  
\bibitem{Arean:2013mta} 
  D.~Arean, A.~Farahi, L.~A.~Pando Zayas, I.~S.~Landea and A.~Scardicchio,
  ``Holographic superconductor with disorder,''
  Phys.\ Rev.\ D {\bf 89}, no. 10, 106003 (2014)
  [arXiv:1308.1920 [hep-th]].
  
\bibitem{Zeng:2013yoa} 
  H.~B.~Zeng,
  ``Possible Anderson localization in a holographic superconductor,''
  Phys.\ Rev.\ D {\bf 88}, no. 12, 126004 (2013)
  [arXiv:1310.5753 [hep-th]].
  
\bibitem{Arean:2014oaa} 
  D.~Arean, A.~Farahi, L.~A.~Pando Zayas, I.~S.~Landea and A.~Scardicchio,
  ``Holographic p-wave Superconductor with Disorder,''
  arXiv:1407.7526 [hep-th].
  
\bibitem{Horowitz:2012ky} 
  G.~T.~Horowitz, J.~E.~Santos and D.~Tong,
  ``Optical Conductivity with Holographic Lattices,''
  JHEP {\bf 1207}, 168 (2012)
  [arXiv:1204.0519 [hep-th]].

\bibitem{Hartnoll:2007ih} 
  S.~A.~Hartnoll, P.~K.~Kovtun, M.~Muller and S.~Sachdev,
  ``Theory of the Nernst effect near quantum phase transitions in condensed matter, and in dyonic black holes,''
  Phys.\ Rev.\ B {\bf 76}, 144502 (2007)
  [arXiv:0706.3215 [cond-mat.str-el]].
  
\bibitem{Hartnoll:2008hs} 
  S.~A.~Hartnoll and C.~P.~Herzog,
  ``Impure AdS/CFT correspondence,''
  Phys.\ Rev.\ D {\bf 77}, 106009 (2008)
  [arXiv:0801.1693 [hep-th]].
  
\bibitem{Hartnoll:2012rj} 
  S.~A.~Hartnoll and D.~M.~Hofman,
  ``Locally Critical Resistivities from Umklapp Scattering,''
  Phys.\ Rev.\ Lett.\  {\bf 108}, 241601 (2012)
  [arXiv:1201.3917 [hep-th]].
  
\bibitem{Lucas:2014sba} 
  A.~Lucas and S.~Sachdev,
  ``Conductivity of weakly disordered strange metals: from conformal to hyperscaling-violating regimes,''
  Nucl.\ Phys.\ B {\bf 892}, 239 (2015)
  [arXiv:1411.3331 [hep-th]].
  
\bibitem{Lucas:2015vna} 
  A.~Lucas,
  ``Conductivity of a strange metal: from holography to memory functions,''
  JHEP {\bf 1503}, 071 (2015)
  [arXiv:1501.05656 [hep-th]].
  
  
  
  \end{thebibliography}
\end{document}